\documentclass[
twocolumn,
hf,
]{ceurart}

\usepackage{lmodern}
\newtheorem{theorem}{Theorem}
\usepackage{bm}
\usepackage{multirow}
\usepackage{amsmath}
\usepackage{mathtools}
\usepackage[inline]{enumitem}
\usepackage{graphicx}
\usepackage{booktabs}
\usepackage{pifont}
\usepackage{mdframed}
\usepackage{dashrule}
\usepackage{caption}
\usepackage{subcaption}
\usepackage{diagbox}
\usepackage{spverbatim}
\usepackage{algorithm,algpseudocode}

\algnewcommand{\LineComment}[1]{\State \(\triangleright\) #1}

\newcommand{\vect}[1]{\bm{#1}}
\DeclarePairedDelimiterX{\infdivx}[2]{(}{)}{%
  #1\;\delimsize\|\;#2%
}
\DeclarePairedDelimiter{\norm}{\lVert}{\rVert}

\DeclarePairedDelimiter{\share}{[}{]}
\DeclarePairedDelimiter{\round}{\lfloor}{\rceil}
\newcommand{\parties}{\ensuremath{\mathcal{P}}}
\DeclareMathOperator{\reveal}{\mathbf{Reveal}}

\algblock{ParFor}{EndParFor}
\algnewcommand\algorithmicparfor{\textbf{parallel for}}
\algnewcommand\algorithmicpardo{\textbf{do}}
\algnewcommand\algorithmicendparfor{\textbf{end\ parallel for}}
\algrenewtext{ParFor}[1]{\algorithmicparfor\ #1\ \algorithmicpardo}
\algrenewtext{EndParFor}{\algorithmicendparfor}

\begin{document}

\copyrightyear{2024}
\copyrightclause{Copyright for this paper by its authors.
  Use permitted under Creative Commons License Attribution 4.0
  International (CC BY 4.0).}

\conference{The IJCAI-2024 AISafety Workshop}

\title{Low-Latency Privacy-Preserving Deep Learning Design via Secure MPC}



\author[1]{Ke Lin}[%
orcid=0009-0002-5376-7881,
email=leonard.keilin@gmail.com,
]
\address[1]{Tsinghua University, 30 Shuangqing Rd., Haidian District, Beijing, 100084, China}


\author[1]{Yasir Glani}[%
orcid=0000-0003-0060-4771,
email=yasirglani@gmail.com,
]

\author[1]{Ping Luo}[%
orcid=0000-0001-6171-3811,
email=luop@tsinghua.edu.cn,
]
\cormark[1]

\cortext[1]{Corresponding author.}





\begin{abstract}
Secure multi-party computation (MPC) facilitates privacy-preserving computation between multiple parties without leaking private information.
While most secure deep learning techniques utilize MPC operations to achieve feasible privacy-preserving machine learning on downstream tasks, the overhead of the computation and communication still hampers their practical application.
This work proposes a low-latency secret-sharing-based MPC design that reduces unnecessary communication rounds during the execution of MPC protocols.
We also present a method for improving the computation of commonly used nonlinear functions in deep learning by integrating multivariate multiplication and coalescing different packets into one to maximize network utilization.
Our experimental results indicate that our method is effective in a variety of settings, with a speedup in communication latency of $10\sim20\%$.
\end{abstract}


\maketitle

\section{Introduction}
Secure multi-party computation (MPC)~\cite{damgard2012multiparty,keller2020mpspdz} enables parties to compute securely over their private data without revealing the data to each other. Secure MPC offers privacy-preserving property, which makes it suitable for most privacy-sensitive domains, such as medical research and finance. 
Upon the development of deep learning techniques, the ability to capture important information from large datasets of neural models raises concerns regarding the surveillance of individuals~\cite{liu2021privacy}.
In this case, the prospects of secure MPC demonstrate its application to secure machine learning and deep learning. 
While MPC-based deep learning frameworks have achieved significant performance in general scenarios, most works suffer from the limitations caused by 
\begin{enumerate*}
    \item \emph{network communication} due to the nature of exchanging intermediate information during MPC execution,
    \item \emph{excessive computation} introduced by complex MPC protocols.
\end{enumerate*}
Since the computation of MPC protocols is largely determined by their sophisticated design, optimizing the protocol itself would seem to be difficult and infeasible. Thus, some studies~\cite{lu2021polymath} are concerned with improving the communication stage of MPC protocols to make them more practical.
In this paper, we present an approach to reduce the communication latency of the MPC protocol through optimized multivariate multiplication.

In general, privacy-preserving deep learning frameworks usually adopt secret-sharing-based techniques to avoid extensive computational overheads~\cite{knott2021crypten,tan2021cryptGPU,li2023mpcformer,Wagh2020FalconHM}. Consequently, secret-sharing-based methods require multiple exchanges of intermediate results to achieve collaborative MPC operations.
As these MPC techniques are based on linear computations, such as addition and multiplication, modern deep learning techniques that inherently rely on linear algebra benefit significantly from them.
Considering the heavy dependency on linear operations, our research aims to reduce unnecessary communication rounds following \cite{krips2022arithmetic} during the execution of MPC protocols.

Our main contributions are as follows:
\begin{itemize}
    \item We propose a low-latency secret-sharing-based method for computing multivariate multiplications and univariate polynomials using network communication that is efficient and reduces unnecessary communication rounds on the fly.
    \item We improve the computation of nonlinear functions by integrating the proposed multivariate multiplication and coalescing different packets into one single packet to maximize network utilization.
    \item We conducted experiments to evaluate the effectiveness of our method in the context of models with varying sizes, networks with different latency and bandwidth, the accuracy of downstream classification tasks, and the number of participants involved. The results indicate an overall improvement of $10\sim20\%$ in communication latency.
\end{itemize}

\section{Background}
\subsection{Arithmetic Secret Sharing Based Scheme}
Our setup is primarily focused on arithmetic operations, so we represent all inputs and intermediate results in terms of linear secret sharing between $n$ parties, especially in the context of additive secret-sharing schemes.

Apart from the general $(n,t)$-Shamir secret sharing scheme~\cite{shamir1979share}, which relies on the degree-$t$ polynomials over $n$ parties, we adopt the simple arithmetic secret sharing scheme based on $(n,0)$-Shamir secret sharing. 
In other words, we share a scalar value $x\in \mathbb{Z}/Q\mathbb{Z}$ across $n$ parties $\parties$, where $\mathbb{Z}/Q\mathbb{Z}$ denotes a ring with $Q$ elements, following the notations of \cite{knott2021crypten}. 
The sharing of $x$ is defined as $\share{x}=\{\share{x}_p\}_{p\in\parties}$, where $\share{x}_p$ is the party $p$'s share of $x$. The ground-truth value $x$ could be reconstructed from the sum of the shares of each party, i.e. $x=\sum_{p\in\parties} \share{x}_p$.

When parties wish to share a value $x$, they generate a pseudorandom zero-share that sums to 0. The party that possesses the value adds $x$ to their share in secret. To represent floating-point numbers, we adopt a fixed-point encoding to encode any floating-point number $x_F$ into a fixed-point representation, $x$. Alternatively, we consider that each $x$ is the result of multiplying a floating-point number $x_F$ by a scaling factor $B=2^L$ and rounding to the nearest integer, i.e. $x=\round{B x_F}$. Here $L$ is the precision of the fixed-point encoding. To decode a ground-truth floating-point value $x_F$ from $x$, we compute as follows: $x_F\approx x/B$.

\subsection{Arithmetic Secret Sharing Based MPC}
\label{sec:bg_ass_MPC}
It is noteworthy that arithmetic secret shares are homomorphic and can be used to implement secure MPC, especially in the context of linear computation in most cases.

\paragraph{Addition.} 
The sum of two secret shared values $\share{x}$ and $\share{y}$ could be directly computed as $\share{z}=\share{x}+\share{y}$, where each party $p\in\parties$ computes $\share{z}_p=\share{x}_p+\share{y}_p$ without multi-party communications.

\paragraph{Multiplication.} 
Two secret shared values $\share{x}$ and $\share{y}$ are multiplied using a random Beaver triple~\cite{beaver1992efficient} generated by the Trusted Third Party (TTP): a triplet $(\share{a},\share{b},\share{ab})$. It should be noted that the Beaver triple could be shared in advance by each party. 
The parties first calculate $\share{\epsilon}=\share{x}-\share{a}$ and $\share{\delta}=\share{y}-\share{b}$. In this way, the $\share{\epsilon}$ and $\share{\delta}$ are then revealed to all parties (denoted as $\reveal(\cdot)$) without compromising information since the ground-truth values $a,b$ remain unknown to each party except for the TTP. 
The final results could be computed as $\share{xy}=\share{c}+\epsilon\share{b}+\share{a}\delta+\epsilon\delta$.
Algorithm \ref{alg:beaver_multiplication} illustrates the multiplication using Beaver triples.

\paragraph{Linear functions.}
It is possible to implement functions that consist of linear operations by combining additions and multiplications. Common operations in deep learning, such as element-wise product and convolution, are allowed in a linear paradigm.

\paragraph{Nonlinear functions.}
Due to the inherent infeasibility of nonlinear functions in the standard arithmetic secret-sharing scheme, most works use approximation methods to simulate the outcome of nonlinear functions. In particular, Taylor Expansion, Newton-Rhapson, and Householder methods are commonly used to approximate nonlinear functions using only linear operations. 
All reciprocal functions, exponential functions, loss functions, kernel functions, and other useful functions in deep learning are calculated this way, for example.

\begin{algorithm}
\caption{Beaver Multiplication $\mathbf{Mul}(\share{x}, \share{y})$}
\label{alg:beaver_multiplication}
\begin{algorithmic}[1]
\Require Secret-shared inputs $\share{x},\share{y}$, Beaver triple ($\share{a},\share{b},\share{ab}$).
\Ensure $\share{xy}$.

\LineComment{Compute masked values}
\State $\share{\epsilon}\gets \share{x-a} = \share{x} - \share{a}$
\State $\share{\delta}\gets \share{y-b} = \share{y} - \share{b}$
\LineComment{Reveal $\epsilon$ and $\delta$ through one-round communications}
\State $\epsilon\gets \reveal(\share{\epsilon})$
\State $\delta\gets \reveal(\share{\delta})$
\State \Return $\epsilon\delta + \epsilon\share{b} + \share{a}\delta + \share{ab}$
\end{algorithmic}
\end{algorithm}

\subsection{Notations}
This section summarizes the notations used throughout this work. We denote $\share{x}$ as a secret sharing of $x$. $\reveal(\share{x})$ means that the ground-truth value $x$ is revealed to every party involved in the computation through one-round communications. Since most linear operations are also applicable to element-wise operations and matrix operations, $x$ can also represent a vector, matrix, or even a tensor if there is no confusion and ambiguity.

\section{Related Work}
To achieve communication-efficient MPC, various approaches have been developed to optimize the communication rounds and the throughput of communication. 
\citet{ishai2000randomizing} proposes a new representation of polynomials for round-efficient secure computation, dividing high-degree polynomials into multiple low-degree polynomials that are easy to solve. 
\citet{mohassel2006efficient} performs operations directly on polynomials, such as polynomial multiplication and division.
\citet{dachman2011secure} improves the evaluation of multivariate polynomials with different variables being held as private inputs by each party.
Then, \citet{lu2021polymath} proposes an efficient method for evaluating high-degree polynomials with arbitrary numbers of variables.
While the current research has focused on improving the calculation of polynomials, our study aims to develop a communication-efficient and effective MPC system for use in modern deep learning frameworks by leveraging arithmetic tuples computation from \citet{krips2022arithmetic}. This system is not confined to only computing polynomials within finite rings, as seen in previous studies.

In recent years, several deep learning frameworks that preserve privacy have emerged to enable the secure inference of neural network models. 
\citet{Wagh2020FalconHM} implements a maliciously secure 3-party MPC protocol from SecureNN~\cite{wagh2018securenn} and ABY\textsuperscript{3}~\cite{mohassel2018aby3}.
\citet{knott2021crypten} provides flexible machine-learning APIs with a rich set of functions for secure deep learning.
\citet{li2023mpcformer} presents a fast and performant MPC Transformer inference framework designed to be privacy-preserving.
Our low-latency linear MPC implementation is built on top of \citeauthor{knott2021crypten}'s \texttt{CrypTen} framework and provides a significant improvement in communication latency.

\section{Methodology}
\subsection{Multivariate Multiplication}
\label{sec:multi_mul}
Since Beaver triples illustrate how to multiply two variables with pre-shared triplets, a classic multiplication between multiple variables, such as $\share{xyz}$, requires several rounds of binary multiplication, i.e. $\mathbf{Mul}(\mathbf{Mul}(\share{x}, \share{y}), \share{z})$. 
This naive implementation, however, introduces additional communication rounds during the on-the-fly $\reveal$ process. In general, a $n$-ary multiplication requires $n-1$ rounds of communication.

To reduce the communication rounds involved in the multivariate multiplications, the basic binary Beaver triple is extended into a general $n$-ary Beaver triple. This results in only \emph{one round} of communication required throughout the entire process. 

Assume the $n$ inputs could be represented as $\{\share{x_i}\}_{i=1}^n$. The precomputation and preshared information required by the extended protocol is $\{\mathcal{A}_i\}_{i=1}^{n}$. 
Here $\mathcal{A}_1 := \{\share{a_j}\}_{j=1}^{n}$ is defined as the set of $n$ auxiliary shared values used to blind the the inputs $\{\share{x_i}\}_{i=1}^n$, which is also similar to the Beaver's idea. Then $\mathcal{A}_i (i\geq 2)$ is defined as the set of shared degree-$i$ cross-terms consisting of the variables in $\mathcal{A}_1$.
For example, $\mathcal{A}_2$ could be defined as $\mathcal{A}_2 := \{\share{a_i a_j}\mid\ i\neq j \wedge 1\leq i,j \leq n\}$, and $\mathcal{A}_3 := \{\share{a_i a_j a_k}\mid\ i\neq j \neq k \wedge 1\leq i,j,k \leq n\}$, and so on. Similar to the construction of $\share{\epsilon}$ and $\share{\delta}$ in Section~\ref{sec:bg_ass_MPC}, we define the difference between the inputs and the masks as $\share{\delta_i} := \share{x_i} - \share{a_i}$. The secret-shared $\share{\delta_i}$ is then made public across all parties without leakage to the ground-truth value of both $\share{x_i}$ and $\share{a_i}$. 
The improvement of our method originates from the following equation:
\begin{equation}
\label{eqa:raw_prod}
\begin{aligned}
\prod_{i=1}^n x_i &= \prod_{i=1}^n (\delta_i + a_i) \\
&= \prod \delta_i + \sum_{i} \delta_i \frac{\prod a_m}{a_i} + \sum_{i,j,i\neq j} \delta_i\delta_j \frac{\prod a_m}{a_i a_j} \\
&\quad+ \dots + \prod a_i.
\end{aligned}
\end{equation}
Here we informally use the fractional representation, such as $\frac{\prod a_m}{a_i}$, to denote the products of all the terms \emph{except for} certain ones. Note that this fractional form does not involve any actual division. Also, each secret-shared term of $\share{\frac{\prod a_m}{a_i\dots a_j}}$ could be found in the auxiliary sets $\{\mathcal{A}_i\}_{i=1}^n$, which is preshared across all parties.

Adaptation of Equation~\ref{eqa:raw_prod} in secret-sharing scheme is as follows:
\begin{equation}
\label{eqa:ss_prod}
\begin{aligned}
\share{\prod_{i=1}^n x_i} &= \prod \delta_i + \sum_{i} \delta_i \share{\frac{\prod a_m}{a_i}} + \sum_{i,j,i\neq j} \delta_i\delta_j \share{\frac{\prod a_m}{a_i a_j}} \\
&\quad+\dots + \share{\prod a_i}.
\end{aligned}
\end{equation}
Since Equation~\ref{eqa:ss_prod} is linear to the secret-sharing terms, all communications could be conducted in parallel, i.e. in a single round of communications. In this case, we could simply reveal all the secret-sharing terms in $\{\mathcal{A}_i\}_{i=1}^n$ and compute the sharing of final results in constant complexity. The protocol is formally described in Algorithm~\ref{alg:multivariate_multiplication}.

\begin{algorithm}
\caption{Multivariate Beaver Multiplication of $n$ inputs $\mathbf{Mul}(\share{x_1}, \share{x_2}, \dots, \share{x_n})$}
\label{alg:multivariate_multiplication}
\begin{algorithmic}[1]
\Require Secret-shared inputs $\{\share{x_i}\}_{i=1}^n$, auxiliary sets $\{\mathcal{A}_i\}_{i=1}^n$.
\Ensure $\share{\prod x_i}$.

\LineComment{Compute masked values}
\For{$i\in [1,n]$}
    \State $\share{\delta_i}\gets \share{x_i - a_i} = \share{x_i} - \share{a_i}$
\EndFor
\LineComment{Reveal $\delta_i$ through one-round communications}
\ParFor{$i\in [1,n]$}
    \State $\delta_i \gets \reveal(\share{\delta_i})$
\EndParFor
\LineComment{Compute results using preshared $\{\mathcal{A}_i\}_{i=1}^n$}
\State \Return $\prod \delta_i + \sum_{i} \delta_i \share{\frac{\prod a_m}{a_i}} + \sum_{i,j,i\neq j} \delta_i\delta_j \share{\frac{\prod a_m}{a_i a_j}} + \dots + \share{\prod a_i}$
\end{algorithmic}
\end{algorithm}

The total rounds of communications are indeed reduced from $n-1$ to constant $1$ when a regular $n$-ary multiplication is performed, but the overall size of communication data increases from linear to exponential. In a na\"ive implementation, the data size of $n$-ary multiplication is only $3(n-1)$ for a total transmission of $n-1$ Beaver triples. As opposed to a multivariate implementation, it is $2^n - 1$ to transmit the auxiliary sets $\{\mathcal{A}_i\}_{i=1}^n$. Therefore, in practice, there is a trade-off between communication latency and throughput.

\subsection{Univariate Polynomials}
The formal form of univariate polynomials is defined as $P(x)=\sum_{i=0}^{n} b_i x^i$, where $b_i$ refers to the coefficients of the degree-$i$ term. The use of univariate polynomials enables efficient evaluation and manipulation of polynomial expressions. 
According to \citet{damgard2006unconditionally}, we can compute all required $\share{x^i}$ in parallel using multivariate multiplications, then multiply them with corresponding plaintext coefficients. Despite its benefits, this trick has the disadvantage of exponentially increasing the size of transmitted data, which becomes unbearable when the exponent exceeds 5.

This method can be implemented in practice by computing a tuple of base terms and then multiplying the tuple by a certain term iteratively, as in the exponentiating by squaring method or the fast modulo algorithm. 
In other words, a tuple $\vect{g}=(1,x,\dots,x^{m-1})$ of size $\norm{\vect{g}}=m$ could be multiplied by $x^{\norm{\vect{g}}}$ repeatedly to iterate all the $x^i$ terms.
The overview of the implementation of univariate polynomials is described in Algorithm~\ref{alg:univariate_poly}. Note that $\vect{b}_{s:e}$ is the subvector of $\vect{b}$ from position $s$ to $e$.

\begin{algorithm}
\caption{Univariate Polynomial $\mathbf{Poly}(\share{x}, \vect{b})$}
\label{alg:univariate_poly}
\begin{algorithmic}[1]
\Require Secret-shared input $\share{x}$, coefficients $\vect{b}=(b_0,b_1,\dots,b_n)$, base terms size $\norm{\vect{g}}$.
\Ensure $\sum_{i=0}^{n} b_i\share{x^i}$.

\LineComment{Construct base terms}
\ParFor{$i\in [1,\norm{\vect{g}}]$}
    \State $\share{x^i}\gets \mathbf{Mul}(\share{x},\dots,\share{x})$ \Comment{multiplied by $\share{x}$ of $i$ times}
\EndParFor
\State $\vect{g}\gets (1,\share{x},\dots,\share{x^{\norm{\vect{g}}-1}})$

\LineComment{Iteratively exponentiating}
\State $t\gets 0$
\For{$i\in[0,\lfloor\frac{n}{\norm{g}}\rfloor - 1]$}
    \State $\text{s}\gets i\cdot\norm{\vect{g}}$
    \State $\text{e}\gets (i+1)\cdot\norm{\vect{g}}$
    \State $t\gets t + \vect{b}_{s:e}\cdot \vect{g}$
    \LineComment{Vectorized Beaver Multiplication}
    \State $\vect{g}\gets \share{x^{\norm{\vect{g}}}} \cdot \vect{g}$
\EndFor
\State \Return $t$
\end{algorithmic}
\end{algorithm}

\subsection{Nonlinear Approximations}
\label{sec:nonlinear_approx}
In this section, we take one step further to optimize the commonly used nonlinear functions by leveraging the property of parallelization of our proposed multivariate multiplication.

\paragraph{Exponentiation.}
Since exponential functions grow in geometrical speed, approximations based on series expansion generally suffer from a significant reduction in accuracy since we do not know the exact value of the input.
Consequently, we resort to the naive iterative approximation, which is capable of utilizing multivariate multiplication effectively:
\begin{displaymath}
    e^{\share{x}} = \lim_{n\rightarrow \infty} \left(1 + \frac{\share{x}}{d^n}\right)^{d^n}.
\end{displaymath}
During each iteration, the $d$-th power of the previous result is calculated. With increasing iteration rounds $n$, the answer would become closer to the actual results.

\paragraph{Logarithm.}
The calculation of logarithms relies on the higher-order iterative methods for a better convergence, i.e. Householder methods on $y=\ln x$:
\begin{displaymath}
\begin{aligned}
    \share{h_n} &= 1 - \share{x} e^{-\share{y_n}} \\
    \share{y_{n+1}} &= \share{y_n} - \sum_{k=1}^{\infty} \frac{1}{k}\share{h_n^k}
\end{aligned}
\end{displaymath}
Note that the implementation of logarithm is the combination of exponentiation and univariate polynomials. The degree of the polynomials determines the precision of the output.

\paragraph{Reciprocal.}
The reciprocal function $y=\frac{1}{x}$ is calculated using the Newton-Raphson method with an initial guess $y_0$:
\begin{displaymath}
\begin{aligned}
    \share{y_{n+1}} &= \share{y_n}(2 - \share{x}\share{y_n}) \\
    &= 2\share{y_n} - \share{x}\share{y_n}\share{y_n}
\end{aligned}
\end{displaymath}

\paragraph{Trigonometry.}
Trigonometric functions could be treated as the special case of exponentiation with $d=2$. The sine and cosine functions are calculated in the field of complex numbers:
\begin{displaymath}
\begin{aligned}
\share{\sin{x}}&=\mathrm{Im}(\share{e^{ix}}) \\
\share{\cos{x}}&=\mathrm{Re}(\share{e^{ix}})
\end{aligned}
\end{displaymath}

Using the above-mentioned nonlinear functions, we can calculate most of the existing loss functions in deep learning, such as the \texttt{sigmoid}, \texttt{tanh}, and cross-entropy functions. Various other common nonlinear functions, such as the \texttt{softmax} function and kernel function, can also be calculated using exponential and reciprocal functions.

\subsection{Communication Coalescing}
The key to achieving low-latency secret-sharing computation is to reduce the total number of rounds of communications among different parties. 
While we introduce a latency-friendly implementation of basic math operations, other kinds of communications, such as precision checking, still require an additional but \emph{independent} communication round.

In general, the communication involved in multiple math operations could be abstracted as a communication graph, or strictly, as a communication tree. Accordingly, we observe some independent communications that do not affect downstream results can be deferred and combined into one single round of communication.
This process is referred to as \emph{communication coalescing}, and it eliminates unnecessary rounds of communication and improves the utilization of network bandwidth.

\section{Security Analysis}
The correctness of the multivariate multiplication is trivial based on the observation in Equation~\ref{eqa:raw_prod} and \ref{eqa:ss_prod}. As univariate polynomials are implemented using the same method as the extended fast modulo algorithm, their effectiveness could also be demonstrated by the correctness and security of multivariate multiplication. 
Coalescing mechanisms only alter the order of communication rounds without modifying the payload, which is also reliable and secure.

Multivariate computations are similarly secure as traditional Beaver multiplications under semi-honest conditions. It is intuitively obvious that since $a_i$ is chosen at random by TTP, the $\delta_i=x_i - a_i$ value is indistinguishable from a random number.
Consequently, the disclosure of $\share{\delta_i}$ does not reveal any critical information regarding $x_i$. This assumption holds even if multiple parties, except for the TTP, collude.

\begin{table*}[t]
\caption{Communication Latency and Data with Different Network Settings. Latency and data are measured in milliseconds and MiBs.}
\label{tab:main}
\centering
\scalebox{0.8}{
\begin{tabular}{cc|c|cc|cccccc}
\toprule
\multirow{2}{*}{\textbf{Model}} & \multirow{2}{*}{\textbf{Dataset}} & \multirow{2}{*}{\textbf{$t_\text{comp}$}} & \multicolumn{2}{c|}{\textbf{Data Size}} & \multicolumn{2}{c}{\textbf{$t_\text{comm}$ ($N_\text{low}$)}} & \multicolumn{2}{c}{\textbf{$t_\text{comm}$ ($N_\text{med}$)}} & \multicolumn{2}{c}{\textbf{$t_\text{comm}$ ($N_\text{high}$)}} \\
\cmidrule(lr){4-5} \cmidrule(lr){6-7} \cmidrule(lr){8-9} \cmidrule(lr){10-11}
 &  &  & \textbf{Na\"ive} & \textbf{Ours} & \textbf{Na\"ive} & \textbf{Ours} & \textbf{Na\"ive} & \textbf{Ours} & \textbf{Na\"ive} & \textbf{Ours} \\
\midrule
LinearSVC & MNIST & 0.032 & 0.022 & 0.024 & 0.082 & 0.059 & 0.604 & 0.568 & 4.339 & 4.026 \\
LeNet & CIFAR-10 & 0.900 & 38.562 & 41.647 & 1.373 & 1.182 & 7.673 & 6.983 & 53.010 & 47.868 \\
ResNet-18 & ImageNet & 110.447 & 11571.294 & 12612.711 & 219.541 & 185.293 & 1681.685 & 1407.570 & \textasciitilde 11~760 & \textasciitilde 9~870 \\
Transformer & Sentiment140 & 7.824 & 771.787 & 893.421 & 12.190 & 9.715 & 78.624 & 61.877 & 559.645 & 421.413 \\
\bottomrule
\end{tabular}
}
\end{table*}

To clarify the security of multivariate multiplication formally, we denote $\share{x}_{p}$ as the secret share of $x$ for party $p\in\mathcal{P}$. The global equations of the multivariate system are as follows:
\begin{equation}
\begin{aligned}
    \sum_{p\in\mathcal{P}} \share{x_i}_p &= x_i \\
    \sum_{p\in\mathcal{P}} \share{a_i}_p &= a_i \\
    \sum_{p\in\mathcal{P}} \share{a_i a_j}_p &= a_i a_j \\
    \dots&\dots \\
    \sum_{p\in\mathcal{P}} \share{a_1\dots a_n} &=  a_1\dots a_n \\
    \share{x_i}_p - \share{a_i}_p &= \share{\delta_i}_p
\end{aligned}
\end{equation}
with known $\share{\delta_i}_p$ for every $p\in\mathcal{P}$ to each party. From each party's view, these $2^n + 2n - 1$ equations have $\Theta(2^n\norm{\mathcal{P}})$ unknown variables. This indicates the difficulty in determining the exact value of $x_i$, as shown in \cite{couteau2019complexity}.

A party's \texttt{view} represents all the values it can obtain during its execution. Then the following theorem holds:
\begin{theorem}
Let $\{x_i'\}$ and $\{x_i''\}$ be random values. The distribution of the view of each party is identical when $x_i=x_i'$ or $x_i=x_i''$.
\end{theorem}
This guarantees the security of multivariate multiplication by ensuring the indistinguishability between the random distribution and the view's distribution.

\section{Experiments}
\subsection{Experimental Setup}
As part of our proposed methodology, we use \texttt{CrypTen}~\cite{knott2021crypten} as the basic MPC deep learning framework, which has already provided na\"ive implementations of secret-sharing-based computations. In most of our experiments, we use 3-party MPC on CPUs.
Additionally, we allow a maximum of $4$-ary multiplication as stated in Section~\ref{sec:multi_mul}, and we set $d=3$ for exponentiation and $k=8$ for logarithm as described in Section~\ref{sec:nonlinear_approx}.

To measure the performance, we perform several experiments with deep learning models with different sizes:
\begin{enumerate*}[label=(\alph*)]
    \item Linear Support Vector Classification (LinearSVC) with L2 penalty;
    \item LeNet~\cite{lecun1998neural} with shallow convolutional and linear layers along with ReLU activation functions;
    \item ResNet-18 model~\cite{he2016deep} with multiple convolutional, linear, pooling, and activation layers;
    \item Transformer Encoder model~\cite{vaswani2017attention} with a single multi-head attention layer and BatchNorm~\cite{ioffe2015batchnorm} in place of LayerNorm~\cite{ba2016layernorm}.
\end{enumerate*}
We employ several datasets for classification tasks with appropriate adaption to specific models, including MNIST~\cite{lecun1998neural}, CIFAR-10~\cite{krizhevsky2009learning}, ImageNet\cite{deng2009imagenet}, and Sentiment140~\cite{go2009twitter} datasets. 

Each of our experiments is conducted in a simulated multi-node environment using Docker. TTP is conducted in an independent environment separate from the normal parties. To manually simulate different network environments concerning bandwidth and latency, we utilize the \texttt{docker-tc} tool to adjust the docker network settings accordingly.

\subsection{Metrics}
To provide a comprehensive evaluation of our proposed method, we adopt metrics from a variety of perspectives.
\begin{itemize}
    \item $t_\text{comp}$: The computational time cost for evaluating a single data sample in one round.
    \item $t_\text{comm}$: The time cost of communication associated with evaluating a single data sample in one round.
    \item \textbf{Size of Transmission Data}: The size of transmitted network packets when evaluating a single data sample in one round.
    \item \textbf{Accuracy}: The classification accuracy when evaluated on a particular dataset.
\end{itemize}


\subsection{Latency \& Throughput}
To assess the efficiency of our proposed method, we simulate networks with different network latencies:
\begin{enumerate*}[label=(\alph*)]
    \item network $N_\text{low}$ with 0.1ms latency,
    \item network $N_\text{med}$ with 5ms latency,
    \item and network $N_\text{high}$ with 40ms latency.
\end{enumerate*}
All of these networks have a bandwidth of 1Gbps. Our simulated multi-node settings include 3 nodes with an additional TTP by default.

As shown in Table~\ref{tab:main}, the computation cost of each model is negligible in medium and high latency network settings in comparison to the communication cost. Therefore, we will focus only on the communication costs associated with our proposed method.

Compared to the na\"ive method implemented by \texttt{CrypTen}, our method illustrated in Section~\ref{sec:multi_mul} remains close since it does not introduce a substantial amount of additional communication payload if the maximum number of input variables is set appropriately. 
For instance, a 3-ary or 4-ary multiplication would not produce a significant increase in the total size of communications.

It is noteworthy that our proposed method reduces the communication cost in every network setting as compared to the na\"ive implementation of MPC. Overall, we achieve an improvement of $10\sim 20\%$, which shows significant enhancement in the performance of high-latency environments for practical purposes.

\begin{figure}[t]
  \centering
  \begin{subfigure}[t]{0.49\linewidth}
    \centering
    \includegraphics[width=\linewidth]{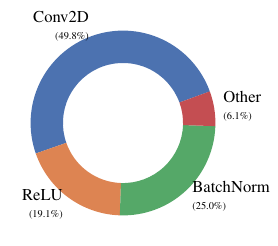}
    \caption{ResNet Basic Block}
  \end{subfigure}
  \begin{subfigure}[t]{0.49\linewidth}
    \centering
    \includegraphics[width=\linewidth]{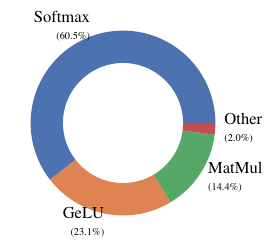}
    \caption{Attention Block}
  \end{subfigure}
  \caption{Communication percentage of different models.}
  \label{fig:occ}
\end{figure}

Furthermore, we observe that our proposed method behaves differently with neural models with different architectures. 
Figure~\ref{fig:occ} illustrates the communication occupation percentage of ResNet basic blocks and Attention blocks.
As can be seen, the attention mechanism is constrained by its communication bottleneck in \texttt{Softmax} operation, while CNN is constrained by its communication via convolutional operations. Considering that our method makes an improved optimization for nonlinear functions, attention-based models show a significant improvement in latency, with almost a 25\% improvement. Additionally, this explains the limited improvement of only $8\sim15\%$ in traditional machine-learning models and CNN-based models.

\begin{table}[t]
\caption{Classification Accuracy using Different Methods.}
\label{tab:accuracy}
\centering
\scalebox{0.88}{
\begin{tabular}{ccccc}
\toprule
\multirow{2}{*}{\textbf{Model}} & \multirow{2}{*}{\textbf{Dataset}} & \multicolumn{3}{c}{\textbf{Accuracy (\%)}} \\
\cmidrule(lr){3-5}
 &  & \textbf{Origin} & \textbf{Na\"ive} & \textbf{Ours} \\
\midrule
LinearSVC & MNIST & 100.00 & 100.00 & 100.00 \\
LeNet & CIFAR-10 & 100.00 & 100.00 & 100.00 \\
ResNet-18 & ImageNet & 69.30 & 61.58 & 60.10 \\
Transformer & Sentiment140 & 59.87 & 58.55 & 57.74 \\
\bottomrule
\end{tabular}
}
\end{table}

\subsection{Evaluation}
In this section, we examine the side effects and factors associated with the basic settings, such as the downstream tasks' accuracy, the number of parties involved, and the trade-off between network latency and bandwidth.

To evaluate the drop in accuracy, we compare our method with both the original baseline and the na\"ive implementation without a low-latency design. 
Figure~\ref{tab:accuracy} shows that, in relatively small scenarios, both the na\"ive implementation and our methods are capable of achieving perfect performance as the baselines.
Nevertheless, both MPC-based implementations obtain lower accuracy in complex scenarios than the baseline, while our methods perform slightly worse than the na\"ive implementation. 
We hypothesize that the multivariate multiplication introduces additional precision requirements, which in turn reduces accuracy.

The throughput and latency of MPC-based methods are also affected by the number of parties involved in the computation. 
From Figure 2, it can be seen that the communication data size of both methods increases linearly as the number of parties involved increases.
There is, however, a tendency for the latency to be worse when there are more parties involved.

\begin{figure}[t]
  \centering
  \begin{subfigure}[t]{0.8\linewidth}
    \centering
    \includegraphics[width=\linewidth]{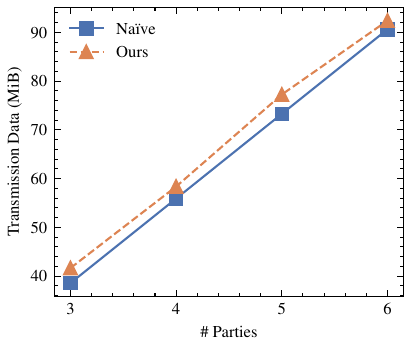}
    \caption{Size of Trans. Data w.r.t \# Parties}
  \end{subfigure}
  \begin{subfigure}[t]{0.8\linewidth}
    \centering
    \includegraphics[width=\linewidth]{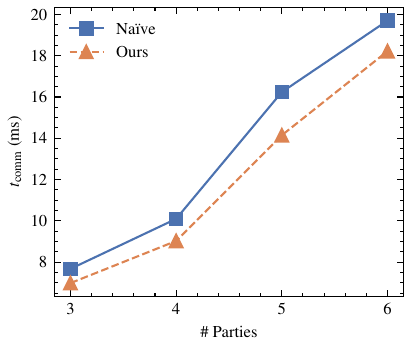}
    \caption{Latency w.r.t \# Parties}
  \end{subfigure}
  \caption{Transimission data and latency of na\"ive and our proposed methods when a different number of parties are involved. The experiment is conducted using the LeNet model on CIFAR-10 using a medium latency network.}
  \label{fig:party}
\end{figure}

\begin{figure}[t]
    \centering
    \includegraphics[width=0.9\linewidth]{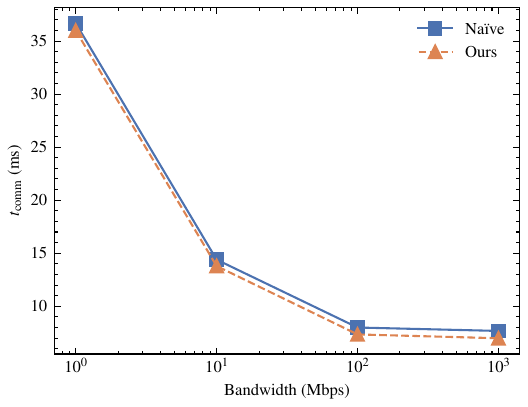}
    \caption{Latency of na\"ive and our proposed methods concerning different network bandwidth. The experiment is conducted using the LeNet model on CIFAR-10 using a medium latency network.}
    \label{fig:bandwidth}
\end{figure}

Moreover, Figure~\ref{fig:bandwidth} illustrates how network bandwidth affects communication costs. 
When sufficient bandwidth is available, our method can still optimize the network latency. 
It is important to note, however, that when the bandwidth becomes the bottleneck, our method would not be any more effective in reducing the overall costs of communication. This indicates that bandwidth remains an important factor in a multi-node MPC setting, especially as the number of nodes in use grows.

\section{Discussion}
Since the proposed multivariate multiplication is based on a finite ring, it is likely to have precision issues that lead to incorrect results. Fortunately, a loss in precision would not significantly affect the overall performance of deep learning, since the loss could be interpreted as random noise and distortion in the input data.

Moreover, our proposed method is only applicable to functions that are based on linear MPC operation. To avoid heavy communications, a modern MPC-based deep learning framework would also involve other protocols, such as Homomorphic Encryption~\cite{gentry2009fully}, Garbled Circuit~\cite{yao1982protocols}, and Function Secret Sharing\cite{boyle2015function}. Though these works may have less communication, our approach could still be seamlessly integrated with the current secret-sharing framework and achieve a latency improvement of \textasciitilde20\% without adding excessive computational workload.

\section{Conclusion}
This study proposes a secret-sharing-based MPC method for enhancing the linear computation required in deep learning through increased communication utilization. By utilizing the multivariate multiplication and communication coalescing mechanisms, we can reduce the number of unnecessary communication rounds during the execution of both linear and nonlinear deep learning functions.
In our experiments, we demonstrate that our proposed methods achieve an overall improvement in latency of $10\sim20\%$ when compared to the na\"ive MPC implementation. Additionally, it indicates that throughput and downstream task performance are comparable to na\"ive implementations, which demonstrate the method's validity and efficiency.
We hope that this work will inspire future improvements in privacy-preserving deep learning techniques and lead to more practical MPC applications.

\begin{acknowledgments}
This work is supported by the National Key R\&D Program of China under grant (No. 2022YFB2703001).
\end{acknowledgments}


\bibliography{mybib}

\begin{thebibliography}{29}
\expandafter\ifx\csname natexlab\endcsname\relax\def\natexlab#1{#1}\fi
\providecommand{\url}[1]{\texttt{#1}}
\providecommand{\href}[2]{#2}
\providecommand{\path}[1]{#1}
\providecommand{\DOIprefix}{doi:}
\providecommand{\ArXivprefix}{arXiv:}
\providecommand{\URLprefix}{URL: }
\providecommand{\Pubmedprefix}{pmid:}
\providecommand{\doi}[1]{\href{http://dx.doi.org/#1}{\path{#1}}}
\providecommand{\Pubmed}[1]{\href{pmid:#1}{\path{#1}}}
\providecommand{\bibinfo}[2]{#2}
\ifx\xfnm\relax \def\xfnm[#1]{\unskip,\space#1}\fi
\bibitem[{Damg{\aa}rd et~al.(2012)Damg{\aa}rd, Pastro, Smart, and Zakarias}]{damgard2012multiparty}
\bibinfo{author}{I.~Damg{\aa}rd}, \bibinfo{author}{V.~Pastro}, \bibinfo{author}{N.~Smart}, \bibinfo{author}{S.~Zakarias},
\newblock \bibinfo{title}{Multiparty computation from somewhat homomorphic encryption},
\newblock in: \bibinfo{editor}{R.~Safavi-Naini}, \bibinfo{editor}{R.~Canetti} (Eds.), \bibinfo{booktitle}{CRYPTO 2012}, \bibinfo{publisher}{Springer Berlin Heidelberg}, \bibinfo{address}{Berlin, Heidelberg}, \bibinfo{year}{2012}, pp. \bibinfo{pages}{643--662}. \DOIprefix\doi{10.1007/978-3-642-32009-5_38}.
\bibitem[{Keller(2020)}]{keller2020mpspdz}
\bibinfo{author}{M.~Keller},
\newblock \bibinfo{title}{Mp-spdz: A versatile framework for multi-party computation},
\newblock in: \bibinfo{booktitle}{Proceedings of the 2020 ACM SIGSAC Conference on Computer and Communications Security}, CCS '20, \bibinfo{publisher}{Association for Computing Machinery}, \bibinfo{address}{New York, NY, USA}, \bibinfo{year}{2020}, p. \bibinfo{pages}{1575–1590}. \DOIprefix\doi{10.1145/3372297.3417872}.
\bibitem[{Liu et~al.(2021)Liu, Xie, Wang, Zou, Xiong, Ying, and Vasilakos}]{liu2021privacy}
\bibinfo{author}{X.~Liu}, \bibinfo{author}{L.~Xie}, \bibinfo{author}{Y.~Wang}, \bibinfo{author}{J.~Zou}, \bibinfo{author}{J.~Xiong}, \bibinfo{author}{Z.~Ying}, \bibinfo{author}{A.~V. Vasilakos},
\newblock \bibinfo{title}{Privacy and security issues in deep learning: A survey},
\newblock \bibinfo{journal}{IEEE Access} \bibinfo{volume}{9} (\bibinfo{year}{2021}) \bibinfo{pages}{4566--4593}. \DOIprefix\doi{10.1109/ACCESS.2020.3045078}.
\bibitem[{Lu et~al.(2021)Lu, Yu, Kate, and Maji}]{lu2021polymath}
\bibinfo{author}{D.~Lu}, \bibinfo{author}{A.~Yu}, \bibinfo{author}{A.~Kate}, \bibinfo{author}{H.~Maji},
\newblock \bibinfo{title}{Polymath: Low-latency mpc via secure polynomial evaluations and its applications},
\newblock \bibinfo{journal}{Proceedings on Privacy Enhancing Technologies} \bibinfo{volume}{2022} (\bibinfo{year}{2021}). \DOIprefix\doi{10.2478/popets-2022-0020}.
\bibitem[{Knott et~al.(2021)Knott, Venkataraman, Hannun, Sengupta, Ibrahim, and van~der Maaten}]{knott2021crypten}
\bibinfo{author}{B.~Knott}, \bibinfo{author}{S.~Venkataraman}, \bibinfo{author}{A.~Hannun}, \bibinfo{author}{S.~Sengupta}, \bibinfo{author}{M.~Ibrahim}, \bibinfo{author}{L.~van~der Maaten},
\newblock \bibinfo{title}{Crypten: Secure multi-party computation meets machine learning},
\newblock \bibinfo{journal}{NIPS} \bibinfo{volume}{34} (\bibinfo{year}{2021}) \bibinfo{pages}{4961--4973}. \DOIprefix\doi{10.48550/arXiv.2109.00984}.
\bibitem[{Tan et~al.(2021)Tan, Knott, Tian, and Wu}]{tan2021cryptGPU}
\bibinfo{author}{S.~Tan}, \bibinfo{author}{B.~Knott}, \bibinfo{author}{Y.~Tian}, \bibinfo{author}{D.~J. Wu},
\newblock \bibinfo{title}{\textsc{CryptGPU}: Fast privacy-preserving machine learning on the gpu},
\newblock in: \bibinfo{booktitle}{{IEEE} {S\&P}}, \bibinfo{year}{2021}. \DOIprefix\doi{10.1109/SP40001.2021.00098}.
\bibitem[{Li et~al.(2023)Li, Wang, Shao, Guo, Xing, and Zhang}]{li2023mpcformer}
\bibinfo{author}{D.~Li}, \bibinfo{author}{H.~Wang}, \bibinfo{author}{R.~Shao}, \bibinfo{author}{H.~Guo}, \bibinfo{author}{E.~Xing}, \bibinfo{author}{H.~Zhang},
\newblock \bibinfo{title}{Mpcformer: fast, performant and private transformer inference with mpc},
\newblock in: \bibinfo{booktitle}{The 11th International Conference on Learning Representations}, \bibinfo{year}{2023}. \DOIprefix\doi{10.48550/arXiv.2211.01452}.
\bibitem[{Wagh et~al.(2020)Wagh, Tople, Benhamouda, Kushilevitz, Mittal, and Rabin}]{Wagh2020FalconHM}
\bibinfo{author}{S.~Wagh}, \bibinfo{author}{S.~Tople}, \bibinfo{author}{F.~Benhamouda}, \bibinfo{author}{E.~Kushilevitz}, \bibinfo{author}{P.~Mittal}, \bibinfo{author}{T.~Rabin},
\newblock \bibinfo{title}{Falcon: Honest-majority maliciously secure framework for private deep learning},
\newblock \bibinfo{journal}{Proceedings on Privacy Enhancing Technologies} \bibinfo{volume}{2021} (\bibinfo{year}{2020}) \bibinfo{pages}{188 -- 208}. \DOIprefix\doi{10.2478/popets-2021-0011}.
\bibitem[{Krips et~al.(2022)Krips, K{\"u}sters, Reisert, and Rivinius}]{krips2022arithmetic}
\bibinfo{author}{T.~Krips}, \bibinfo{author}{R.~K{\"u}sters}, \bibinfo{author}{P.~Reisert}, \bibinfo{author}{M.~Rivinius},
\newblock \bibinfo{title}{Arithmetic tuples for mpc},
\newblock \bibinfo{journal}{Cryptology ePrint Archive}  (\bibinfo{year}{2022}). \URLprefix \url{https://eprint.iacr.org/2022/667}.
\bibitem[{Shamir(1979)}]{shamir1979share}
\bibinfo{author}{A.~Shamir},
\newblock \bibinfo{title}{How to share a secret},
\newblock \bibinfo{journal}{Commun. ACM} \bibinfo{volume}{22} (\bibinfo{year}{1979}) \bibinfo{pages}{612–613}. \DOIprefix\doi{10.1145/359168.359176}.
\bibitem[{Beaver(1992)}]{beaver1992efficient}
\bibinfo{author}{D.~Beaver},
\newblock \bibinfo{title}{Efficient multiparty protocols using circuit randomization},
\newblock in: \bibinfo{editor}{J.~Feigenbaum} (Ed.), \bibinfo{booktitle}{CRYPTO 1991}, \bibinfo{publisher}{Springer Berlin Heidelberg}, \bibinfo{address}{Berlin, Heidelberg}, \bibinfo{year}{1992}, pp. \bibinfo{pages}{420--432}. \DOIprefix\doi{10.1007/3-540-46766-1_34}.
\bibitem[{Ishai and Kushilevitz(2000)}]{ishai2000randomizing}
\bibinfo{author}{Y.~Ishai}, \bibinfo{author}{E.~Kushilevitz},
\newblock \bibinfo{title}{Randomizing polynomials: A new representation with applications to round-efficient secure computation},
\newblock in: \bibinfo{booktitle}{Proceedings 41st Annual Symposium on Foundations of Computer Science}, \bibinfo{year}{2000}, pp. \bibinfo{pages}{294--304}. \DOIprefix\doi{10.1109/SFCS.2000.892118}.
\bibitem[{Mohassel and Franklin(2006)}]{mohassel2006efficient}
\bibinfo{author}{P.~Mohassel}, \bibinfo{author}{M.~Franklin},
\newblock \bibinfo{title}{Efficient polynomial operations in the shared-coefficients setting},
\newblock in: \bibinfo{editor}{M.~Yung}, \bibinfo{editor}{Y.~Dodis}, \bibinfo{editor}{A.~Kiayias}, \bibinfo{editor}{T.~Malkin} (Eds.), \bibinfo{booktitle}{PKC 2006}, \bibinfo{publisher}{Springer Berlin Heidelberg}, \bibinfo{address}{Berlin, Heidelberg}, \bibinfo{year}{2006}, pp. \bibinfo{pages}{44--57}. \DOIprefix\doi{10.1007/11745853_4}.
\bibitem[{Dachman-Soled et~al.(2011)Dachman-Soled, Malkin, Raykova, and Yung}]{dachman2011secure}
\bibinfo{author}{D.~Dachman-Soled}, \bibinfo{author}{T.~Malkin}, \bibinfo{author}{M.~Raykova}, \bibinfo{author}{M.~Yung},
\newblock \bibinfo{title}{Secure efficient multiparty computing of multivariate polynomials and applications},
\newblock in: \bibinfo{editor}{J.~Lopez}, \bibinfo{editor}{G.~Tsudik} (Eds.), \bibinfo{booktitle}{Applied Cryptography and Network Security}, \bibinfo{publisher}{Springer Berlin Heidelberg}, \bibinfo{address}{Berlin, Heidelberg}, \bibinfo{year}{2011}, pp. \bibinfo{pages}{130--146}. \DOIprefix\doi{10.1007/978-3-642-21554-4_8}.
\bibitem[{Wagh et~al.(2018)Wagh, Gupta, and Chandran}]{wagh2018securenn}
\bibinfo{author}{S.~Wagh}, \bibinfo{author}{D.~Gupta}, \bibinfo{author}{N.~Chandran},
\newblock \bibinfo{title}{Securenn: Efficient and private neural network training},
\newblock \bibinfo{journal}{Cryptology ePrint Archive}  (\bibinfo{year}{2018}). \URLprefix \url{https://eprint.iacr.org/2018/442}.
\bibitem[{Mohassel and Rindal(2018)}]{mohassel2018aby3}
\bibinfo{author}{P.~Mohassel}, \bibinfo{author}{P.~Rindal},
\newblock \bibinfo{title}{Aby3: A mixed protocol framework for machine learning},
\newblock in: \bibinfo{booktitle}{Proceedings of the 2018 ACM SIGSAC Conference on Computer and Communications Security}, CCS '18, \bibinfo{publisher}{Association for Computing Machinery}, \bibinfo{address}{New York, NY, USA}, \bibinfo{year}{2018}, p. \bibinfo{pages}{35–52}. \DOIprefix\doi{10.1145/3243734.3243760}.
\bibitem[{Damg{\aa}rd et~al.(2006)Damg{\aa}rd, Fitzi, Kiltz, Nielsen, and Toft}]{damgard2006unconditionally}
\bibinfo{author}{I.~Damg{\aa}rd}, \bibinfo{author}{M.~Fitzi}, \bibinfo{author}{E.~Kiltz}, \bibinfo{author}{J.~B. Nielsen}, \bibinfo{author}{T.~Toft},
\newblock \bibinfo{title}{Unconditionally secure constant-rounds multi-party computation for equality, comparison, bits and exponentiation},
\newblock in: \bibinfo{editor}{S.~Halevi}, \bibinfo{editor}{T.~Rabin} (Eds.), \bibinfo{booktitle}{Theory of Cryptography}, \bibinfo{publisher}{Springer Berlin Heidelberg}, \bibinfo{address}{Berlin, Heidelberg}, \bibinfo{year}{2006}, pp. \bibinfo{pages}{285--304}. \DOIprefix\doi{10.1007/11681878_15}.
\bibitem[{Couteau(2019)}]{couteau2019complexity}
\bibinfo{author}{G.~Couteau},
\newblock \bibinfo{title}{A note on the communication complexity of multiparty computation in the correlated randomness model},
\newblock in: \bibinfo{booktitle}{EUROCRYPT 2019}, \bibinfo{publisher}{Springer-Verlag}, \bibinfo{address}{Berlin, Heidelberg}, \bibinfo{year}{2019}, p. \bibinfo{pages}{473–503}. \DOIprefix\doi{10.1007/978-3-030-17656-3_17}.
\bibitem[{Lecun et~al.(1998)Lecun, Bottou, Bengio, and Haffner}]{lecun1998neural}
\bibinfo{author}{Y.~Lecun}, \bibinfo{author}{L.~Bottou}, \bibinfo{author}{Y.~Bengio}, \bibinfo{author}{P.~Haffner},
\newblock \bibinfo{title}{Gradient-based learning applied to document recognition},
\newblock \bibinfo{journal}{Proceedings of the IEEE} \bibinfo{volume}{86} (\bibinfo{year}{1998}) \bibinfo{pages}{2278--2324}. \DOIprefix\doi{10.1109/5.726791}.
\bibitem[{He et~al.(2016)He, Zhang, Ren, and Sun}]{he2016deep}
\bibinfo{author}{K.~He}, \bibinfo{author}{X.~Zhang}, \bibinfo{author}{S.~Ren}, \bibinfo{author}{J.~Sun},
\newblock \bibinfo{title}{Deep residual learning for image recognition},
\newblock in: \bibinfo{booktitle}{CVPR}, \bibinfo{year}{2016}, pp. \bibinfo{pages}{770--778}. \DOIprefix\doi{10.1109/CVPR.2016.90}.
\bibitem[{Vaswani et~al.(2017)Vaswani, Shazeer, Parmar, Uszkoreit, Jones, Gomez, Kaiser, and Polosukhin}]{vaswani2017attention}
\bibinfo{author}{A.~Vaswani}, \bibinfo{author}{N.~Shazeer}, \bibinfo{author}{N.~Parmar}, \bibinfo{author}{J.~Uszkoreit}, \bibinfo{author}{L.~Jones}, \bibinfo{author}{A.~N. Gomez}, \bibinfo{author}{L.~u. Kaiser}, \bibinfo{author}{I.~Polosukhin},
\newblock \bibinfo{title}{Attention is all you need},
\newblock in: \bibinfo{editor}{I.~Guyon}, \bibinfo{editor}{U.~V. Luxburg}, \bibinfo{editor}{S.~Bengio}, \bibinfo{editor}{H.~Wallach}, \bibinfo{editor}{R.~Fergus}, \bibinfo{editor}{S.~Vishwanathan}, \bibinfo{editor}{R.~Garnett} (Eds.), \bibinfo{booktitle}{NIPS}, volume~\bibinfo{volume}{30}, \bibinfo{publisher}{Curran Associates, Inc.}, \bibinfo{year}{2017}. \DOIprefix\doi{10.5555/3295222.3295349}.
\bibitem[{Ioffe and Szegedy(2015)}]{ioffe2015batchnorm}
\bibinfo{author}{S.~Ioffe}, \bibinfo{author}{C.~Szegedy},
\newblock \bibinfo{title}{Batch normalization: accelerating deep network training by reducing internal covariate shift},
\newblock in: \bibinfo{booktitle}{Proc. 32nd Int. Conf. Machine Learning - Volume 37}, ICML'15, \bibinfo{publisher}{JMLR.org}, \bibinfo{year}{2015}, p. \bibinfo{pages}{448–456}. \DOIprefix\doi{10.5555/3045118.3045167}.
\bibitem[{Ba et~al.(2016)Ba, Kiros, and Hinton}]{ba2016layernorm}
\bibinfo{author}{J.~L. Ba}, \bibinfo{author}{J.~R. Kiros}, \bibinfo{author}{G.~E. Hinton}, \bibinfo{title}{Layer normalization}, \bibinfo{year}{2016}. \href{http://arxiv.org/abs/1607.06450}{{\tt arXiv:1607.06450}}.
\bibitem[{Krizhevsky et~al.(2009)Krizhevsky, Hinton et~al.}]{krizhevsky2009learning}
\bibinfo{author}{A.~Krizhevsky}, \bibinfo{author}{G.~Hinton}, et~al.,
\newblock \bibinfo{title}{Learning multiple layers of features from tiny images}  (\bibinfo{year}{2009}).
\bibitem[{Deng et~al.(2009)Deng, Dong, Socher, Li, Li, and Fei-Fei}]{deng2009imagenet}
\bibinfo{author}{J.~Deng}, \bibinfo{author}{W.~Dong}, \bibinfo{author}{R.~Socher}, \bibinfo{author}{L.-J. Li}, \bibinfo{author}{K.~Li}, \bibinfo{author}{L.~Fei-Fei},
\newblock \bibinfo{title}{Imagenet: A large-scale hierarchical image database},
\newblock in: \bibinfo{booktitle}{CVPR}, \bibinfo{year}{2009}, pp. \bibinfo{pages}{248--255}. \DOIprefix\doi{10.1109/CVPR.2009.5206848}.
\bibitem[{Go et~al.(2009)Go, Bhayani, and Huang}]{go2009twitter}
\bibinfo{author}{A.~Go}, \bibinfo{author}{R.~Bhayani}, \bibinfo{author}{L.~Huang},
\newblock \bibinfo{title}{Twitter sentiment classification using distant supervision},
\newblock \bibinfo{journal}{CS224N project report, Stanford} \bibinfo{volume}{1} (\bibinfo{year}{2009}) \bibinfo{pages}{2009}.
\bibitem[{Gentry(2009)}]{gentry2009fully}
\bibinfo{author}{C.~Gentry},
\newblock \bibinfo{title}{Fully homomorphic encryption using ideal lattices},
\newblock in: \bibinfo{booktitle}{Proceedings of the Forty-First Annual ACM Symposium on Theory of Computing}, STOC '09, \bibinfo{publisher}{Association for Computing Machinery}, \bibinfo{address}{New York, NY, USA}, \bibinfo{year}{2009}, p. \bibinfo{pages}{169–178}. \DOIprefix\doi{10.1145/1536414.1536440}.
\bibitem[{Yao(1982)}]{yao1982protocols}
\bibinfo{author}{A.~C. Yao},
\newblock \bibinfo{title}{Protocols for secure computations},
\newblock in: \bibinfo{booktitle}{23rd Annual Symposium on Foundations of Computer Science (sfcs 1982)}, \bibinfo{year}{1982}, pp. \bibinfo{pages}{160--164}. \DOIprefix\doi{10.1109/SFCS.1982.38}.
\bibitem[{Boyle et~al.(2015)Boyle, Gilboa, and Ishai}]{boyle2015function}
\bibinfo{author}{E.~Boyle}, \bibinfo{author}{N.~Gilboa}, \bibinfo{author}{Y.~Ishai},
\newblock \bibinfo{title}{Function secret sharing},
\newblock in: \bibinfo{editor}{E.~Oswald}, \bibinfo{editor}{M.~Fischlin} (Eds.), \bibinfo{booktitle}{EUROCRYPT 2015}, \bibinfo{publisher}{Springer Berlin Heidelberg}, \bibinfo{address}{Berlin, Heidelberg}, \bibinfo{year}{2015}, pp. \bibinfo{pages}{337--367}. \DOIprefix\doi{10.1007/978-3-662-46803-6_12}.

\end{thebibliography}


\end{document}